\documentclass[aps,prb,reprint,amsmath,amssymb,floatfix,superscriptaddress]{revtex4-2}

\usepackage{soul}
\usepackage{graphicx}
\usepackage{dcolumn}
\usepackage{bm}
\usepackage{siunitx}
\usepackage[T1]{fontenc}
\usepackage{mathptmx}
\usepackage{makecell}
\usepackage{threeparttable}
\usepackage{color}
\usepackage{hyperref}
\usepackage{braket}
\definecolor{hotblue}{RGB}{46,48,146}
\hypersetup{
            colorlinks=true,
            linkcolor=hotblue,
            citecolor=hotblue,
            urlcolor=hotblue,
            }

\begin{document}

\title{{Statistical} limits for {entanglement swapping} with semiconductor entangled photon sources}

\author{Jingzhong Yang}
\thanks{Contributed equally to this work}
\author{Michael Zopf}
\thanks{Contributed equally to this work}
\author{Pengji Li}
\thanks{Contributed equally to this work}
\affiliation{ 
Institut f{\"u}r Festk{\"o}rperphysik, Leibniz Universit{\"a}t Hannover, Appelstra{\ss}e~2, 30167~Hannover, Germany}

\author{Nand Lal Sharma}
\affiliation{Institute for Integrative Nanosciences, Leibniz IFW Dresden, Helmholtzstra{\ss}e~20, 01069~Dresden, Germany}

\author{Weijie Nie}
\affiliation{Institute for Integrative Nanosciences, Leibniz IFW Dresden, Helmholtzstra{\ss}e~20, 01069~Dresden, Germany}

\author{Frederik Benthin}

\author{Tom Fandrich}

\author{Eddy Patrick Rugeramigabo}
\affiliation{ 
Institut f{\"u}r Festk{\"o}rperphysik, Leibniz Universit{\"a}t Hannover, Appelstra{\ss}e~2, 30167~Hannover, Germany}

\author{Caspar Hopfmann}
\affiliation{Institute for Integrative Nanosciences, Leibniz IFW Dresden, Helmholtzstra{\ss}e~20, 01069~Dresden, Germany}

\author{Robert Keil}
\altaffiliation[Present address: ]{Fraunhofer-Institut f{\"u}r Angewandte Festk{\"o}rperphysik (IAF), Tullastra{\ss}e 72, 79108~Freiburg, Germany}
\affiliation{Institute for Integrative Nanosciences, Leibniz IFW Dresden, Helmholtzstra{\ss}e~20, 01069~Dresden, Germany}

\author{Oliver G. Schmidt}
\affiliation{Institute for Integrative Nanosciences, Leibniz IFW Dresden, Helmholtzstra{\ss}e~20, 01069~Dresden, Germany}
\affiliation{Material Systems for Nanoelectronics, Technische Universit{\"a}t Chemnitz, 09107~Chemnitz, Germany}
\affiliation{Nanophysics, Faculty of Physics and W{\"u}rzburg-Dresden Cluster of Excellence ct.qmat, TU Dresden, 01062 Dresden, Germany}

\author{Fei Ding}
\email{fei.ding@fkp.uni-hannover.de}
\affiliation{ 
Institut f{\"u}r Festk{\"o}rperphysik, Leibniz Universit{\"a}t Hannover, Appelstra{\ss}e~2, 30167~Hannover, Germany}
\affiliation{Laboratorium f{\"u}r Nano- und Quantenengineering, Leibniz Universit{\"a}t Hannover, Schneiderberg 39, 30167 Hannover, Germany}

\begin{abstract}
{Semiconductor quantum dots are promising building blocks for quantum communication applications. Although deterministic, efficient, and coherent emission of entangled photons} has been realized, implementing a practical quantum {repeater} remains outstanding. Here we explore the {statistical} limits for {entanglement swapping with} sources of polarization-entangled photons from the commonly used biexciton-exciton cascade. We stress the necessity of tuning the exciton fine structure, and explain why the often observed time evolution of photonic entanglement in quantum dots is not applicable for large quantum networks. {We identify the critical, statistically distributed device parameters for entanglement swapping based on two sources. A numerical model for benchmarking the consequences of device fabrication, dynamic tuning techniques, and statistical effects is developed, in order to bring the realization of semiconductor-based quantum networks one step closer to reality.}

\end{abstract}

\maketitle
\section{Introduction}
Entanglement is a fundamental resource in next-generation quantum technologies such as quantum communication \cite{Zeilinger_1998, Pant_2019, Basso_Basset_2021} or quantum computing~\cite{Zhong1460}. The efficient distribution of entanglement between remote parties is a key-enabling element for the realization of a global quantum internet \cite{Ren_2017,PhysRevA.95.032306}. Photons are considered the best "flying" quantum bits for this goal since they can travel long distances with high resistance to  decoherence from the environment \cite{kwiat2000experimental}. Quantum information is encoded on the light by means of observables with continuous~\cite{zhang2020long} or discrete ~\cite{andersen2015hybrid} eigenvalues. The polarization degree of single photons has been utilized to transfer quantum states via optical fiber~\cite{Rosenberg_2007} or satellite signals~\cite{Ursin_2007}. However, the employed sources of single photons and polarization-entangled photon pairs are based on spontaneous parametric down-conversion (SPDC)~\cite{Bouwmeester1997, PhysRevLett.80.3891}, a probabilistic process comprising a fundamental efficiency limit impeding practical applications \cite{RevModPhys.84.777, scarani2005four}.

Self-assembled semiconductor quantum dots (QDs) have been studied extensively in recent decades and have become promising candidates for the generation of single photons \cite{senellart2017high, wang2019towards}, entangled pairs \cite{stevenson2006semiconductor,liu2019solid}, or linear cluster states \cite{schwartz2016deterministic}. However, distributed quantum networks based on QDs have yet to be demonstrated. Such networks may be realized using quantum repeater schemes~\cite{van2020extending} which, among others, rely on quantum interference between photons from independent sources and entanglement swapping~\cite{chen2017experimental}. Photon states generated by swapping entanglement of photon pairs emitted from a single QD have been shown to violate Bells inequality \cite{PhysRevLett.123.160502}. The individual properties of QDs that impact the success of entanglement swapping schemes have been well understood \cite{rota2020entanglement, schimpf2021quantum, scholl2020crux}. However, in the real-world application of distributed devices, it is impossible to choose the best possible values of each parameter simultaneously, since each parameter shows a statistical distribution in each device. 

Here we show how the statistical distribution of QD properties limits their practical application in {entanglement swapping between distributed nodes.} Two separate dielectric antenna devices are studied, which have been reported to significantly enhance the photon extraction efficiency of QDs emitting at near-infrared \cite{chen2018highly} and telecom  \cite{Yang2020} wavelengths. The distribution of QD properties in the bare wafer and the changes induced by the device fabrication process are investigated. {A suitable pair of QDs from two individual devices is tuned into resonance with each other and characterized for usability in entanglement swapping schemes \mbox{\cite{PhysRevLett.123.160502, PhysRevLett.123.160501}}. Polarization-entangled photon pairs are hereby generated via the biexciton-exciton (XX-X) cascade.} We discuss the necessity of tuning the exciton fine structure of each individual device for network protocols involving {Bell state measurements (BSMs). A numerical model is presented by which the fidelity of  entanglement swapping with two sources is estimated based on the statistically distributed QD parameters in each dielectric antenna device.} The presented findings shed light on the roadmap for optimizing and utilizing semiconductor photon sources in distributed quantum networks.

\section{{Spectral overlap between devices}}
\begin{figure*}[http]
    \centering
    \includegraphics[width=\textwidth]{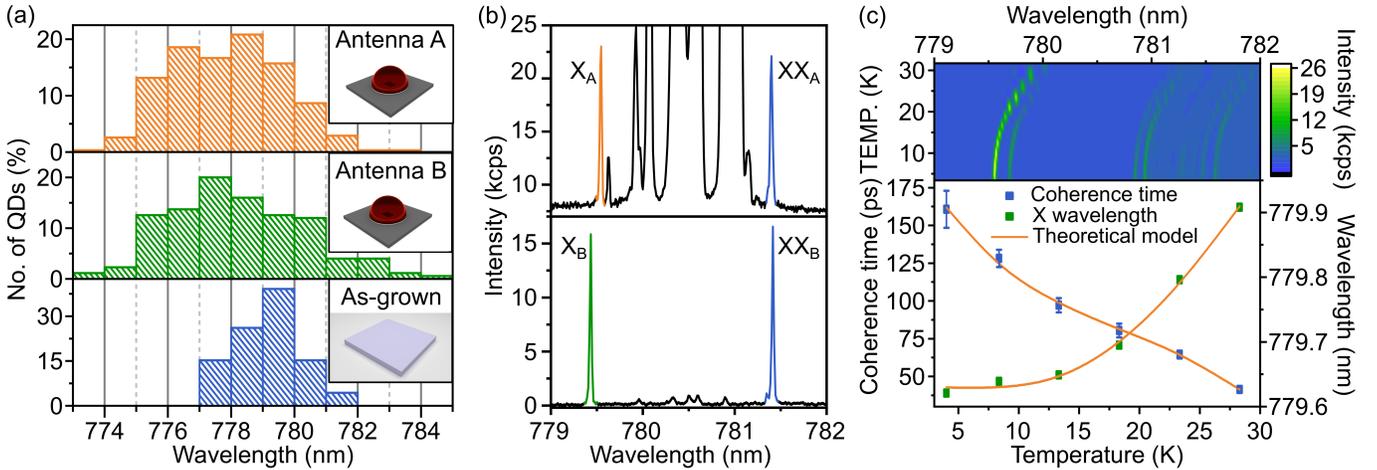}
    \caption{Emission wavelength characteristics and tuning. (a) Statistics of the emission wavelength in fabricated dielectric antenna structures A (orange) and B (green), compared to the as-grown sample (blue), revealing only a slight change after fabrication. {(b) Spectra of two selected QDs with spectral proximity in each dielectric antenna device, excited via two-photon resonant pumping of the biexciton (XX). Different XX binding energies in each QD lead to different spectral distances between the XX and X emissions. The signals between the XX and X peaks represent a residual resonant laser background.} (c) Tuning of QD emission characteristics via temperature, revealing simultaneous changes in emission wavelength, intensity, and coherence time. Photoluminescence spectra (top) and coherence time of the neutral X photons (bottom) under off-resonant excitation for different temperatures.}
    \label{fig:wavelength}
\end{figure*}
{Optical quantum interference lies at the heart of entanglement swapping schemes, requiring mode overlap of indistinguishable photons of the same wavelength for successful BSMs.} In contrast to SPDC sources, whose emission wavelength can be easily adjusted, QDs exhibit discrete spectral lines which differ from dot to dot. Therefore, the emissions have to be tuned into resonance with each other, e.g., via quantum frequency conversion \cite{Weber_2018}, which is, however, limited in efficiency due to coupling and conversion losses as well as higher operation complexity. Another possibility is applying external (e.g., strain or magnetic) fields \cite{chen2016wavelength,Trotta_2016,Bayer_2002}, for which it is important to understand the initial compatibility of two optical devices in wavelength: A too broad distribution of emission wavelengths increases the difficulty of finding similar QDs that can be tuned into resonance. Figure \ref{fig:wavelength}(a) shows the distribution of X emission wavelengths of two dielectric antenna devices containing GaAs/AlGaAs QDs (see details about the samples in the Supplemental Material \cite{supplementary2022_statists}). This virtual strain-free QD system is known for the high level of control over the optical properties, resulting in similar emission spectra and low X fine structure splittings (FSSs) \cite{keil2017solid}. Most of the QDs in the fabricated antennas emit photons at wavelengths between \SI{773}{\nano\metre} and \SI{785}{\nano\metre}. On the one hand, this distribution is blueshifted and broadened compared to the as-grown sample, which can be attributed to different local strain conditions for the QDs after device fabrication. On the other hand, both wavelength distributions are very similar, rendering the fabrication process uniform and offering a certain probability of finding a QD in each device with close-by emission wavelengths. 

{Now, two QDs (labeled A and B) in each dielectric antenna device are chosen with a small difference of emission wavelengths between XX$_A$ and XX$_B$ of \mbox{\SI{15.6}{\pico\meter}}. Their spectral characteristics are illustrated in fig. \mbox{\ref{fig:wavelength}(b)}, obtained by two-photon resonant excitation of the biexciton using $\pi$ pulses.} External tuning techniques then allow for tuning the emission wavelengths of QDs, e.g., via temperature \cite{Rastelli_2007,Benyoucef_2009}, strain \cite{chen2016wavelength,Trotta_2016}, electric fields \cite{Findeis_2001,Zhang2016}, or magnetic fields \cite{Bayer_2002}. 

Applying external fields often results in a mutual change of parameters, as shown in fig. \ref{fig:wavelength}(c). Tuning the temperature from \SI{3.4}{\kelvin} to \SI{31.7}{\kelvin} results in a redshift of the X emission by \SI{0.445}{\nano\meter} \cite{O_Donnell_1991,Holewa_2020}. The temperature-induced change in the band gap typically dose not lead to changes in the X fine structure and lifetime. However, increased charge fluctuations and scattering process with phonons lead to degraded first-order coherence (increased linewidths), setting a limit to the visibility for quantum interference with photons from another device. The coherence time ($T_2$) of the neutral X emission as a function of the temperature was obtained with a Michelson interferometer and is shown. A more than threefold decrease in coherence time is observed when tuning the temperature from \SI{3.4} to \SI{31.4}{\kelvin}. While the wavelength barely changes for low temperatures, it shifts more efficiently for a temperature greater than \SI{15}{\kelvin} [Fig. \ref{fig:wavelength}(c)]. 

\section{{Repeater-relevant device characteristics}}
\begin{figure*}[http]
    \centering
    \includegraphics[width=\textwidth]{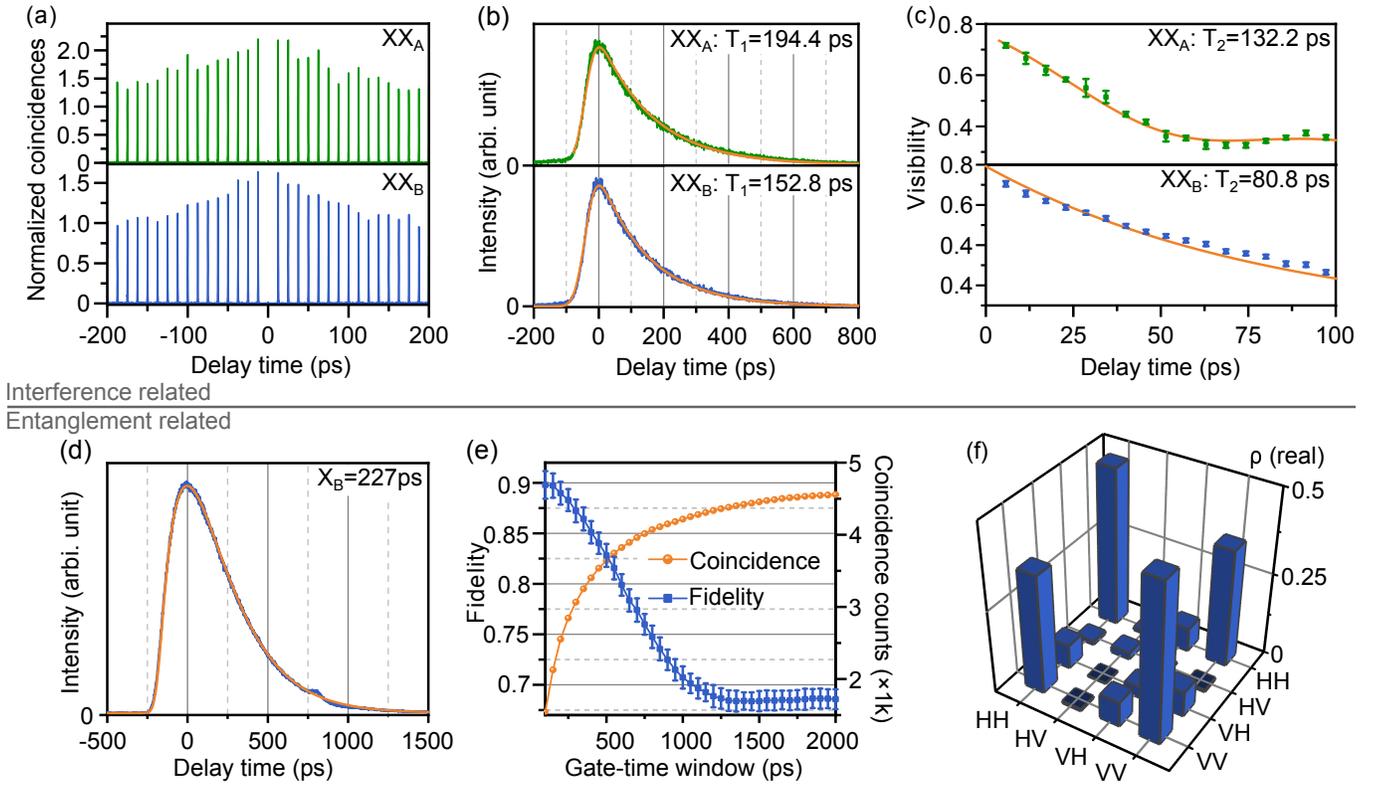}
    \caption{Optical properties from separate devices required for efficient entanglement swapping. (a)--(c) Quantum interference-related properties of the XX photons from two devices (A and B) that have been tuned into resonance via temperature tuning. (a) Second-order autocorrelation measurements for the XX photon streams of both devices, revealing pure single-photon emission as well as blinking. (b) Lifetime measurements and (c) coherence time measurements of both XX photon streams. (d)--(f) Entanglement-related properties of the QD in device B. {(d) Radiative lifetime of X photons.} (e) Entanglement fidelity and raw coincidence counts, obtained via quantum tomography based on polarization-resolved cross-correlation measurements between XX and X emissions, as a function of gate-time window size. (f) Real part of the two-photon density matrix obtained at a gate window size of \SI{500}{\pico\second}.}
    \label{fig:characteristics}
\end{figure*}
{We now investigate the revelent parameters for an entanglement swapping scheme with different sources. The temperature of QD B is now adjusted to \mbox{\SI{10.81}{\kelvin}}, thereby realizing spectral overlap of the XX transitions of QD A and QD B. Due to slight differences in the XX binding energies (in our case, $\approx\SI{0.26}{\milli\electronvolt}$), the X and XX transitions cannot be tuned in resonance with each other simultaneously.} Figures \ref{fig:characteristics}(a)--(c) show the relevant properties for quantum interference, i.e., the BSM in entanglement swapping, and figs. \ref{fig:characteristics}(d)--(f) for entanglement of the initial pairwise entanglement.  Figure \ref{fig:characteristics}(a) illustrates the intensity autocorrelation $g^{(2)}(\tau)$ from the XX streams of devices A and B. At zero time delay, $g_{A}^{2}(0)= g_{B}^{2}(0) \approx 0.03$ is determined, indicating a high single-photon purity leading to negligible degradation on quantum interference \cite{ollivier2021hong} between photons from the two devices. Apart from that, the data are superimposed by a bunching towards zero time delay, giving evidence to blinking. Such emission intermittence \cite{PhysRevB.92.245439} can be explained by residual charges in the QD blocking the resonant excitation of the neutral XX. The "on" fractions are $\SI{46.9(3)}{}\pm\SI{0.42}{\percent}$ and $\SI{51.1(2)}{}\pm\SI{0.83}{\percent}$ for device A and B \cite{supplementary2022_statists}, respectively \cite{nguyen2013photoneutralization,diederichs:tel-01416901}. Blinking \cite{Hopfmann_2021,hopfmann2020deterministic} is detrimental for the efficiency of quantum interference \cite{PhysRevB.96.075430} [in our case, estimated to $\SI{48.9(4)}{\percent}$] and therefore also entanglement swapping. 

The success of quantum interference is furthermore governed by the photon indistinguishability of each photon source, typically determined by $V_{\delta E}=T_2/2T_1$. Figures \ref{fig:characteristics}(b) and 2(c) show the lifetime and coherence time measurements from the XX photons of each device. The indistinguishabilities are therefore estimated to $V_{\delta E}^{(XX_A)}=\SI{34.0(0)}{\percent}$ and $V_{\delta E}^{(XX_B)}=\SI{26.4(4)}{\percent}$, respectively. The limited coherence is usually attributed to dephasing due to charge and spin noise or phonon scattering \cite{zhai2021quantum}. The ratio of the lifetimes of XX and X [i.e., 0.67(3) for QD in device B] is limiting the indistinguishability even further (in our case, to $\sim \SI{15.7}{\percent}$) due to intrinsic dephasing in the cascade emission \cite{huber2013measurement,scholl2020crux,schimpf2021quantum}. A shortening of the XX lifetime, e.g., by the Purcell effect, is therefore beneficial for addressing both discussed points that limit the indistinguishability simultaneously \cite{PhysRevB.92.161302,scholl2020crux,schimpf2021quantum}.

\begin{table*}[ht]
    \caption{\label{tab:parameters} Summary of typical GaAs/AlGaAs QD characteristics relevant in entanglement swapping, together with the used initial values for the model. {(Given uncertainties denote the standard deviation of statistical measurements.)}}
    \begin{threeparttable}
    \begin{ruledtabular}
    \begin{tabular}{ccc}
    \textrm{Parameters}&\textrm{Typical values}&\textrm{\makecell[c]{Initial values \\ for our model}}\\
    \colrule
    Wavelength (X) (nm) & $779.8\pm 1.6$ \cite{keil2017solid} & $777.85\pm 2.19$ \\
    Fine structure (X) (\SI{}{\micro\electronvolt})  & $4.8\pm2.4$ \cite{keil2017solid} & $11\pm6.5$ \\
    X lifetime (ps)  & {$213.75\pm42.75 \tnote{1}$}  & $300 \pm 50$  \\
    XX lifetime (ps) &  {$127.5\pm14.67 \tnote{1}$}  & $150 \pm 25$ \\
    Pure dephasing time (ns) & {$0.41\pm0.16 \tnote{2}$} & $0.5 \pm 0.25$ \\
    $g^{(2)}(0)$ (XX)& $0.02$\cite{keil2017solid},  0.001\cite{liu2019solid} & N.C. \tnote{3}\\
    Spin scattering time (X) (ns) & $15$ \cite{keil2017solid} & N.C. \tnote{3}\\
    Pair collection efficiency ($\SI{}{\percent}$) & 37.2\cite{chen2018highly}, 65.4 \cite{liu2019solid} & N.C. \tnote{3}
    \end{tabular}
        \begin{tablenotes}
        \footnotesize
        \item[1] {\mbox{References \cite{chen2018highly,PhysRevLett.123.160501,keil2017solid,Hopfmann_2021,scholl2020crux,liu2019solid,schimpf2021quantum,Huber2017}}}.
        \item[2] {The statistics of the dephasing time is estimated by the lifetime and linewidth of the charge exciton of 10 QDs in \mbox{\cite{zhai2020low}}}.
        \item[3] {N.C.: not considered in our model.}
    \end{tablenotes}
    \end{ruledtabular}
    \end{threeparttable}
\end{table*}

The experimentally relevant degree of entanglement is determined by the X fine-structure splitting, radiative lifetime, and spin scattering times. In the GaAs/AlGaAs material system, QDs can be found where spin scattering has only a small influence \cite{keil2017solid}. The X lifetime measurement of device B is shown in fig. \ref{fig:characteristics}(d). Although a lower X lifetime would lead to less phase noise in the entangled state, it deteriorates quantum interference visibilities at the same time, and therefore the success of BSM required in the entanglement swapping scheme. The QD in device B exhibits a fine-structure splitting of \SI{4.22(8)}{\micro\electronvolt}. The polarization entanglement is obtained by polarization-resolved cross-correlation measurements and quantum tomography \cite{Altepeter_2005}. Gating of detection events can be applied to enhance the measured degree of entanglement at the expense of coincidence counts \cite{PhysRevLett.123.160502}. This relationship is displayed in fig. \ref{fig:characteristics}(e). The fidelity of the entanglement can be strongly enhanced when the gate-time window size is reduced to \SI{500}{\pico\second}. The corresponding real part of the two-photon density matrix is shown in fig. \ref{fig:characteristics}(f) (the imaginary part is shown in the Supplemental Material \cite{supplementary2022_statists}). However, the coincidence counts also decrease significantly. Time gating can also lead to the improved success of BSMs in entanglement swapping \cite{PhysRevLett.123.160502}, also with the expense of reduced overall coincidences. The efficiency of entanglement swapping, i.e., the four fold coincidence detection rate, is furthermore determined by the source efficiency, including the excitation and collection efficiencies as well as the coupling efficiency into a single-mode fiber which depends on the mode profile of the emission.

{We can eventually identify the most critical, statistically distributed device parameters which affect entanglement swapping with polarization-entangled photon pairs generated via the XX-X cascade from different devices. Typical values for GaAs/AlGaAs QDs found in the literature are listed in Table \mbox{\ref{tab:parameters}}. Together with the experimental data obtained here, we now define initial values for a theoretical model to estimate entanglement swapping performance, as will be described in sec. \mbox{\uppercase\expandafter{\romannumeral5}}.} Although the pure dephasing times may be different for the XX and X emissions, we assume it to be identical here since both transitions are affected by dephasing due to the solid-state environment in a similar way. For actual networks of multiple sources or coupling to quantum memories, the binding energy of the XX also plays a role and is statistically distributed, but is not mentioned in the table above. 

\section{{Necessity of fine structure tuning}}
\begin{figure*}[http]
    \centering
    \includegraphics[width=\textwidth]{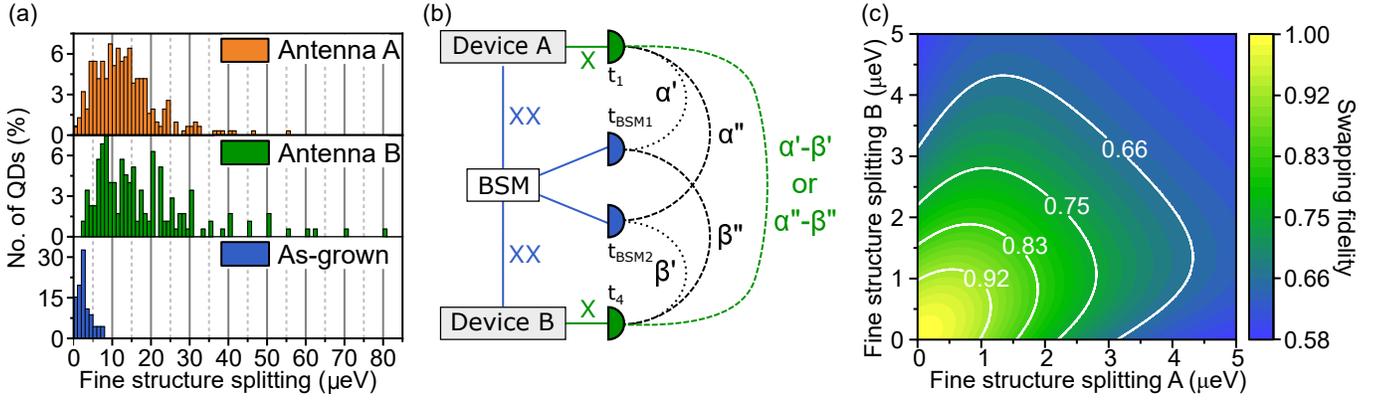}
    \caption{Exciton fine-structure characteristics and influence on entanglement swapping. (a) Statistics of the X fine structure splitting in fabricated dielectric antenna structures A (orange) and B (green), compared to the as-grown sample (blue), revealing a significant broadening of the distribution after fabrication. (b) Scheme of entanglement swapping using QDs with finite exciton fine structure. The phases in the entangled states emitted by devices A and B are not accessible via measuring detection times. Two possible solutions arise ($\alpha^{\prime},\beta^{\prime}$ and $\alpha^{\prime\prime},\beta^{\prime\prime}$), leading to two swapped states in a statistical superposition with the phases $\alpha^{\prime} - \beta^{\prime}$ and $\alpha^{\prime\prime}-\beta^{\prime\prime}$, which cannot be experimentally distinguished. (c) Fidelity of an entanglement swapping scheme with ideal quantum interference at the BSM, taking into account the finite exciton fine structure of two separate QDs and an exciton lifetime of \SI{300}{\pico\second}.}
    \label{fig:fss}
\end{figure*}
For experiments such as entanglement swapping with QD-based light sources, a high swapping fidelity is accomplished only if the sources emit high-fidelity entangled photon pairs themselves. In recent years, it has been shown several times that high degrees of entanglement persist even with larger X fine-structure splittings if one experimentally determines the phase factor for each emission \cite{winik2017demand},
\begin{equation}
\ket{\psi_{12}(t)} = \frac{1}{\sqrt{2}} \left( \ket{H_1 H_2} + e^{- i \frac{S}{\hbar} t} \ket{V_1V_2} \right)
\end{equation}
where the phase factor is determined by the X fine-structure splitting $S$ and the time $t$ that the exciton evolves. The distribution of the X fine-structure splitting undergoes a strong change during fabrication. From fig. \ref{fig:fss}(a), it is clear that the QDs in the as-grown sample have FSSs lower than \SI{10}{\micro\electronvolt}, much smaller than the ones in the dielectric antennas. A large FSS leads to two-photon states that oscillate between two Bell states with a randomized phase \cite{winik2017demand}. This phase can be determined by measuring the detection events of the XX and X photons from each emission and therefore obtaining $t$. However, it is important to note that these types of time-resolved measurements are accompanied by a significant loss in overall efficiency for any protocol that relies on BSMs with photons from different sources \cite{Anderson_2020}, i.e., quantum teleportation or entanglement swapping. This even accounts for experiments using photon pairs emitted by the same source but at different times. Let us consider an emission process of entangled pairs from two separate devices with subsequent detection events in an entanglement swapping scheme as illustrated in fig. \ref{fig:fss}(b). Each emission from device $A$ or $B$ therefore carries its respective phase factor $\alpha = - \frac{S_A}{\hbar} t_A$ or $\beta = - \frac{S_B}{\hbar} t_B$,
\begin{equation}
\begin{split}
    \ket{\psi_{1234}(\alpha,\beta)} = \frac{1}{2} ( \ket{H_1 H_2} + e^{i \alpha} \ket{V_1V_2} ) \ ( \ket{H_3 H_4} + e^{i \beta} \ket{V_3V_4} )
\end{split}
\end{equation}
After projecting photons 2 and 3 on the Bell state $\ket{\Psi^-_{2,3}} = \frac{1}{\sqrt{2}} \left( \ket{H_2 V_3} - \ket{V_2H_3} \right)$, one obtains the un-normalized state,
\begin{equation}
\begin{split}
    \ket{\psi_{1,4}(\alpha,\beta)} &= \braket{\Psi^-_{2,3} | \psi_{1234}(\alpha,\beta)} \\  &= e^{i\beta} \left( \ket{H_1V_4} - e^{i(\alpha-\beta)}  \ket{V_1H_4} \right)
\end{split}
\end{equation}
which means that similar to the case of entangled pair emission from a single device, the final (swapped) state is oscillating between two Bell states, here with the phase $\alpha - \beta$. In order to utilize this state, this phase has to be an experimentally accessible parameter. {However, the wave functions of indistinguishable photons incident on a nonpolarizing beam splitter obey the symmetric exchange condition \mbox{\cite{Holbrow_2002,Weihs2001}}. The BSM therefore removes the ‘‘which path’’ information}, making it impossible to determine from which source each photon originated. Two possible combinations of phase factors are possible:
\begin{equation}
\alpha^\prime =  - \frac{S_A}{\hbar} \left(t_1 - t_{BSM1}\right) \ \ \ \text{and} \ \ \  \beta^\prime =  - \frac{S_B}{\hbar} \left( t_4 - t_{BSM2} \right)
\end{equation}
or 
\begin{equation}
    \alpha^{\prime\prime} =  - \frac{S_A}{\hbar} \left( t_1 - t_{BSM2} \right) \ \ \ \text{and} \ \ \   \beta^{\prime\prime} =  - \frac{S_B}{\hbar} \left( t_4 - t_{BSM1} \right)
\end{equation}
where the indices $i=1,4$ in $t_i$ correspond to the photons described in the state $\ket{\psi_{1234}}$, and the $i=(BSM1),(BSM2)$ correspond to the photons detected at the two respective detectors of a BSM. Independent of the present fine-structure splittings, it is impossible to experimentally determine which of the two cases occurs and therefore which of the two states,
\begin{equation}
    \ket{\psi_{1,4}(\alpha^\prime,\beta^\prime)} \ \ \neq \ \ \ket{\psi_{1,4}(\alpha^{\prime\prime},\beta^{\prime\prime})}
\end{equation}
is present. The final state is therefore not a coherent, but a statistical superposition of the two possible states $\ket{\psi_{1,4}(\alpha^\prime,\beta^\prime)}$ and $\ket{\psi_{1,4}(\alpha^{\prime\prime},\beta^{\prime\prime})}$. The entanglement of the swapped state only persists in the case of low X fine-structure splittings and radiative decay times or the case of $t_{BSM,1} = t_{BSM,2}$. The latter case is undesirable since it strongly lowers the efficiency of the BSM by discarding many detection events $t_{BSM,1} \neq t_{BSM,2}$ within the coincidence window. A finite FSS in separate devices significantly lowers the fidelity of the two-photon states obtained in an entanglement swapping scheme. This is illustrated in fig. \ref{fig:fss}(c) under the assumption of  ideal quantum interference at the BSM for clarification of the theoretical concept. {In reality, quantum interference is degraded by dephasing processes, the intrinsic dynamics of the three-level cascade decay \mbox{\cite{huber2013measurement,scholl2020crux,schimpf2021quantum}}, as well as the  frequency detuning caused by the FSS, resulting in the reduction of photon indistinguishability \mbox{\cite{Neuwirth_2021}}}. The latter two effects can be mitigated by selectively reducing the XX lifetime. Nevertheless, based on the above discussion, a scalable application of QD-based entangled photon sources must rely on tuning the fine structure to remove the state oscillation. This can be achieved, e.g., via frequency shifting optical setups \cite{Wang_2010,Fognini_2018} or external fields such as anisotropic strain. Once such tuning is realized with high efficiency, it allows for strong pairwise photonic entanglement from individual QDs which can then be applied in quantum networks.

\section{Predictions for a real-world quantum network}
\begin{figure*}[http]
    \centering
    \includegraphics[width=\textwidth]{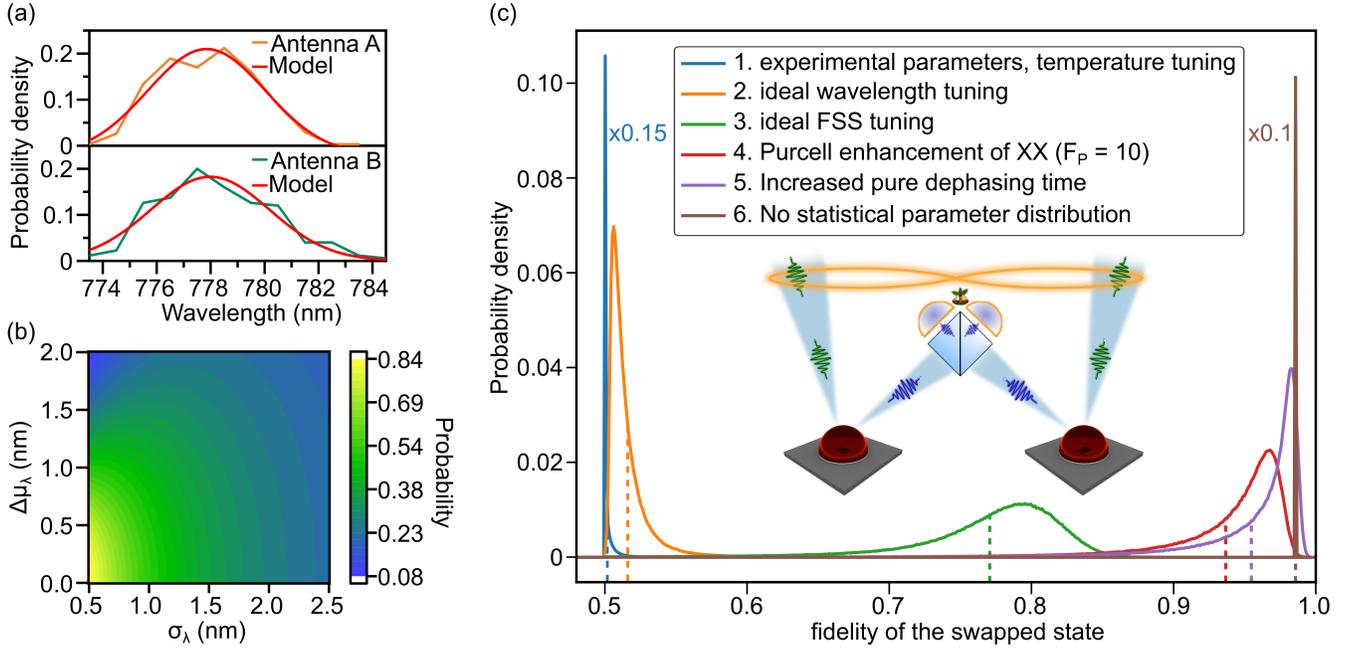}
    \caption{Estimated device performance in a quantum network based on entanglement swapping. The properties of individual QDs are defined by probability density distributions as shown in (a) for the emission wavelengths from devices A and B. The experimental data are fitted with a Gaussian curve. Other properties (FSS, $T_1$, $T_2$, not shown) are modeled by truncated Gaussian distributions.
    (b) Probability for tuning the emission wavelength of two randomly selected QDs from separate devices into resonance, assuming a wavelength tuning range of $\delta_{\lambda} = $\SI{1}{\nano\meter}. The emission wavelength standard deviation $\sigma_{\lambda}$ is set equal for both devices, while the expected value has a difference of $\Delta\mu_{\lambda}$. (c) Probability density of the entanglement fidelity after swapping entanglement between photon pairs emitted by separate devices. The semi-analytical model incorporates a random sampling of experimentally reasonable QD property distributions from both devices. One million samples are used to obtain the probability density. The blue line (curve 1) shows the fidelity distribution for the experimentally accessed parameters in this work. Possible improvements in the fidelity distribution by consecutively adding tuning mechanisms and optimizing device properties are shown in curves 2 to 5. Curve 6 shows the case with removed statistical parameter distribution (for an explicit display, the heights of curves 1 and 6 are scaled down by a factor of $\times 0.15$ and $\times 0.1$, respectively).}
    \label{fig:theory}
\end{figure*}

Even though prominent results have been obtained in optimizing individual parameters of single quantum dots, it is still unclear how to repeatably produce usable devices, due to the statistical performance from device to device. To quantitatively investigate the requirements of the parameters from QD-based photon sources for quantum networks, we now develop a theoretical model which takes into account the statistical effect on the fidelity of photons that result from entanglement swapping. We start by considering two independent sources, which are tunable in wavelength to target the spectral overlap of the emission wavelengths. To estimate the magnitude of the overlap, first the probability density of finding a QD with a specific wavelength in device A and B is obtained (based on Fig. \ref{fig:wavelength}) and shown in Fig. \ref{fig:theory}(a). Under the assumption that the probability density of the emission wavelengths obeys a standard distribution, we can describe it by
\begin{equation}
    p(\lambda)=\frac{1}{\sqrt{2\pi}\sigma} \cdot e^{-\frac{(\lambda-\mu)^2}{2\sigma^2} } 
\end{equation}
where $p(\lambda)$ is the probability density, and $\mu, \sigma$ are the expected and standard deviation values of the distribution. By fitting the data, it is possible to extract the corresponding values \cite{supplementary2022_statists}. Next, the tuning range for the employed wavelength tuning of each device $\delta_{\lambda}$ is defined. Now we can analytically calculate the probability of tuning two devices into resonance \cite{supplementary2022_statists}. Figure \ref{fig:theory}(b) shows the result, assuming an equal $\sigma_{\lambda}$ in the devices, but different expected values with a difference of $\Delta\mu_{\lambda}$. Here, an exemplary wavelength tuning range of \SI{1}{\nano\meter} for each device is assumed. The colormap shows that high probability can be obtained with a decreasing expected value difference $\Delta\mu_{\lambda}$ and narrower wavelength standard deviation $\sigma_{\lambda}$. Furthermore, for higher $\Delta\mu_{\lambda}$ and a fixed tuning range, it is beneficial to have $\sigma_{\lambda}$ that is not too small, in order to increase the chance of tuning remote devices into resonance. 
The model is now extended further, in order to obtain the fidelity of the swapped photon pairs based on the statistically distributed parameters in the devices. The exact nature of the distribution is not clear for every parameter and can be adapted for each specific QD material and fabricated device. Here we assume that the probability density for all parameters (e.g., X fine structure splitting) follows a (truncated) Gaussian distribution. This assumption models the fine-structure well, as can be seen in fig. \ref{fig:theory}(a). The rest of the parameters were approximated by reasonable experimental values (see Table~\ref{tab:parameters} and Supplemental Material \cite{supplementary2022_statists}) based on the presented experiments and the existing literature.

The model is based on a numerical Monte Carlo approach. First, a random set of parameters is drawn for each parameter from devices A and B. A tuning mechanism can now be considered which e.g., tunes the wavelengths from the different devices into resonance, potentially also affecting the other parameters. Then, the swapping fidelity is analytically calculated {under the assumption of ideal projection of entangled states in the BSM, which has been demonstrated with high quality in practice \mbox{\cite{PhysRevLett.123.160502}}}. This process of random sampling is repeated one-million times. The resulting fidelities of the swapped state are then normalized to display a probability density. Figure \ref{fig:theory}(c) shows the resulting probability distribution for obtaining a certain entanglement swapping fidelity for several considered cases. The integration of the probability density within a certain fidelity range then can yield the total probability for obtaining a fidelity within this range in a swapping experiment. Here, we consider a BSM with polarizing beam splitters, which is known to result in a lower limit of 0.5 in the swapping fidelity \cite{PhysRevLett.123.160502}. This coincides with the upper fidelity limit for classical two-photon states. Since we do not consider spin scattering or limited single-photon purity here, the swapped fidelities will not fall below 0.5. The blue curve (1) shows the probability density considering the experimental parameters obtained in this work and assuming temperature tuning. It is clear that entanglement swapping is mostly not successful, leading to fidelities close to 0.5. The reason is the limited tuning range via temperature and the simultaneous decrease in photon coherence. For the orange curve (2), we now assume that all QDs from both devices can be tuned into resonance with each other (e.g., by strain tuning) without affecting other properties. This leads to a visible increase of the entanglement fidelity. However, the values are still located mostly below 0.6. The reason is the X fine structure in the devices. Therefore, we assume an additional tuning mechanism for the fine structure (e.g., anisotropic strain tuning) with a magnitude of $\SI{50}{\micro\electronvolt}$. The resulting green curve (3) now shows fidelities between 0.75 and 0.85. If, furthermore, a selective Purcell enhancement of the XX emission with a Purcell factor of $F_P = 10$ is added \cite{wang2019demand}, the red curve (4) arises. Another significant increase in entanglement swapping fidelity is the result, with fidelities ranging between 0.85 and 0.97. To increase the probability of a high swapping fidelity, even more, the pure dephasing times can be increased to \SI{4}{\nano\second}. The result is shown by the violet curve (5), exhibiting fidelities between 0.9 and 0.99. Curve (6) shows the comparison to the expected probability density if the parameters would not be statistically distributed at all. Therefore, the statistical effect of the QD properties on entanglement swapping is clear by comparing curves (5) and (6). Only a very small chance exists that two QDs can be found such that the swapped fidelity is actually above those of curve (6). Due to the distribution of properties, the swapped entanglement deteriorates in almost all cases. How much lower the fidelity becomes is directly related to the relative uncertainties in the distributed parameters. It is worth mentioning that the same effect would be seen when using differently fabricated sources, e.g., based on circular Bragg gratings or diode structures. High fidelities can therefore only be obtained by either accepting very low sample yield (choosing only the best QDs) or by reducing the parameter uncertainties.

The developed model can be optimized further by implementing the following effects, leading to an even higher accuracy when estimating the fidelity of the swapped photonic states. X spin scattering times can be considered, though they are expected to be long enough to not strongly deteriorate the entanglement of the emitted photon pairs \cite{keil2017solid}. Furthermore, the single-photon purity can be considered, which can be close to unity in GaAs/AlGaAs QDs \cite{Schweickert_2018}. The detrimental effect of the fine structure on the quantum interference at the BSM can be implemented. However, when the XX lifetime is decreased via asymmetric Purcell enhancement, this effect becomes less dominant due to the increased radiatively limited XX linewidths. Also, the total efficiency of the entanglement swapping operation can be included, based on the efficiency of each QD in each device and possibly applied postselection of single-photon counting events.

Now the question has to be raised as to where the advantage in using QD-based entangled photon sources lies. For SPDC sources, a match of emission wavelength between independent sources can be easily achieved by tuning the pump laser or phase-matching conditions. In the low-efficiency limit, this can lead to the generation of highly indistinguishable photons and photon pairs with high fidelities \cite{zhang2016generation}, which is the reason why these types of sources are used in various entanglement-related quantum applications. In contrast, QDs allow for a deterministic emission of photon pairs without any physical limits. However, to really offer an advantage over SPDC sources, QD-based sources must show better total efficiencies while offering the same possible fidelities of swapped photon pairs. For QD-based sources to achieve these fidelities, material growth and device fabrication have to be controlled to an extent that allows for highly narrow distributions in optical properties. Tuning methods of emission wavelength and fine structure are required. Simultaneously, QDs may need to be embedded in devices suppressing blinking and increasing coherence and photon indistinguishability. 

\section{Conclusion}
We present insights on the scalability of semiconductor-enabled quantum photonic networks based on experimental observations and a developed numerical model. Considering QD-based dielectric optical antennas in an entanglement swapping scheme serves as an example for unraveling the challenges for employing semiconductor QDs in applications such as quantum repeaters. {Two QDs from two individual devices are tuned into resonance with each other by temperature tuning, and the properties of the single- and entangled-photon characteristics are evaluated.} We showed that the exciton fine structure is detrimental for scalable networks and therefore requires tuning. The statistical distribution of parameters (e.g., emission wavelengths and fine structures) in two separate devices is studied, and the influence on the achievable entanglement swapping fidelities is investigated. 

Different tuning techniques are contemplated, revealing that the desired tuning of one parameter is often accompanied by a simultaneous tuning of others. Anisotropic strain tuning is most promising for independent tuning of independent parameters (i.e., wavelength and fine structure). Further challenges such as blinking and limited coherence can be addressed, e.g., by implementing QD devices with dynamic charge tuning \cite{zhai2020low}. A selective Purcell enhancement of the XX emission can be obtained using circular Bragg gratings \cite{liu2019solid}. However, the combination of all three techniques is technologically very demanding. In addition, since a large parameter space exists for generating polarization-entangled photon pairs from the XX-X cascade, not all important parameters can be tuned deterministically. A random sampling of several properties is thus unavoidable, posing intensified requirements to the material growth in order to obtain scalable semiconductor devices. The deviation of device parameters may necessitate entanglement purification, which uses up available resources in a quantum network and therefore may relativize the key advantage of source efficiency of QD-based sources. Therefore, the level of control over the parameter statistics directly translates to a potential advantage of QD-based sources in a quantum network. A possible alternative is to generate photonic entanglement with a significantly larger number of photons per individual QD first (e.g., cluster states) \cite{schwartz2016deterministic}, before interfacing with photons from other sources.

This discussion and the numerical prediction can easily be extended to other types of solid-state quantum light emitters or quantum memories with photon interfaces (comprising entanglement generation via photon interference). The model incorporates well-known analytical approaches for the independent physical problems (entangled two-photon states for individual QDs and subsequent entanglement swapping), whereas a numerical solution based on random sampling serves to evaluate the influence of statistical distributions of parameters in the chosen material system. This distribution eventually leads to the degradation of remote entanglement, necessitating time gating or correction protocols such as entanglement purification, which reduce the overall quantum network efficiency. It is therefore a key challenge for the scalable implementation of solid-state quantum light sources. 

\begin{acknowledgments}
The authors gratefully acknowledge the funding by the German Federal Ministry of Education and Research (BMBF) within the project Q.Link.X (16KIS0869) and QR.X (16KISQ015), the European Research Council (QD-NOMS GA715770) and the Deutsche Forschungsgemeinschaft (DFG, German Research Foundation) under Germany's Excellence Strategy (EXC- 2123) QuantumFrontiers (390837967).
\end{acknowledgments}
\appendix
\nocite{}
\bibliography{reference}
\end{document}